\documentclass{svjour3}                     
\smartqed  
\usepackage{graphicx}
\usepackage{mathptmx}      
\begin{document}

\title{Spatial averaging and a non-Gaussianity. }

\author{N.A. Koshelev  }

\institute{ Ulyanovsk State Pedagogical University, 100 years
V.I.Lenin's Birthday Sq.,4, Ulyanovsk 432700, Russia \\
              \email{koshna71@inbox.ru}  }

\date{Received: date / Accepted: date}

\maketitle

\begin{abstract}
The spatial averaging used  for the splitting of the local  scale
factor on the homogeneous background and small inhomogeneous
perturbation leads to a non-local relationship between locally and
globally defined comoving curvature perturbations. We study this
relationship within a quasi-homogeneous, nearly spatially flat
domain of the Universe. It is shown that, on scales larger than
the size of the observed patch, the Fourier components of the
locally defined comoving curvature perturbation are suppressed.
We have also shown that the statistical properties of local and
global comoving curvature perturbations are coincide on a small
scale. Several examples are discussed in detail.

\end{abstract}

\section{Introduction.}
\label{intro}

The consideration of cosmological inhomogeneities within
non-linear perturbation theory has attracted much attention in
recent years. Investigation of non-Gaussian features  of
cosmological perturbations gives new opportunities to test
inflationary models, theoretical models of thermal reheating of
the Universe. It also provides a possibility to get new
cosmological constraints on a number of high energy physics
theories.

Currently, the non-linear perturbation theory is rapidly
developing. The gradient expansion approach
\cite{SB},\cite{9411040},\cite{9905064} led to the generalization
of the $\delta N$ formalism
\cite{Starobinsky},\cite{Sasaki_Stewart},\cite{Sasaki_Tanaka} to
the non-linear case  \cite{0411220},\cite{0504045} that gives a
handy tool for an estimation and calculation of non-Gaussianities.
The  quantum field theory methods are increasingly penetrating
into the cosmological perturbation theory. Non-perturbative
approaches are under development. For example, there are such
techniques as renormalized cosmological perturbation theory
\cite{0509418}, \cite{0509419} and the time renormalization group
method \cite{0806.0971}. However, there are still many unresolved
theoretical issues.

The non-linear generalization of the curvature perturbation
implies the need to take into account the finiteness of the
observable patch of the Universe in determining the local comoving
curvature perturbation.  This makes itself evident in the fact
that the observer identifies the perturbation as deviation from
the spatial average over the observable patch. The consequences of
such definition probably related to the problem of infrared
divergences in the calculation of $n$-point correlation functions
by perturbation theory \cite{0707.0361} are still not well
understood. The potential consequences have been considered
recently \cite{1301.3128},\cite{1303.3549} in the context of a
landscape picture of the Universe. In this picture, observed
statistical properties of perturbations depend on the position of
the observable region. A landscape Universe arises in eternal
inflation \cite{Linde},\cite{Steinhardt}, and some of its
aspects were discussed in connection with curvaton scenario
\cite{0511736}, \cite{1012.0549}. However, it is worth noting that
several questionable approximations have been applied in the
papers \cite{1301.3128}, \cite{1303.3549}.

This paper is organized as follow. In Section \ref{ch2}, we
consider a quasi-homogeneous domain of the Universe and its
observable patch. We introduce globally and locally defined
comoving curvature perturbations ($\zeta (x^\mu)$ and
$\zeta_S(x^\mu)$, correspondingly) and briefly describe the
relationship between them. The exact relations between the Fourier
components of $\zeta $ and $\zeta_S$ are considered  in Section
\ref{ch3}.  We show that, on a scale much smaller than the size of
an observable patch, the values of globally and locally defined
comoving curvature perturbations are coincided with good accuracy.
Section  \ref{ch4} serves to clarify some issues of application
obtained in Section \ref{ch3} equations. In Section  \ref{ch5}, we
discuss the obtained discrepancy with some claims of papers
\cite{1301.3128}, \cite{1303.3549}. We conclude the paper in
Section \ref{ch6}.

\section{The comoving curvature perturbation.}
\label{ch2}

Let's consider a domain of the Universe, which can be described by
nearly homogeneous and spatially flat metric. The comoving
curvature perturbation $\zeta (\mathbf{x})$ is defined by the
local scale factor \cite{0411220}
\begin{equation}
\label{locala}\tilde{a}(\mathbf{x},t) = a(t) e^{\zeta (\mathbf{x})}.
\end{equation}
Equivalently, one can write
\begin{equation}
\label{localzeta}\zeta (\mathbf{x})=\ln \left(\frac{\tilde{a}}{a}\right).
\end{equation}
The ambiguity of this definition is eliminated by condition
\begin{equation}
\label{constr}\langle\zeta (\mathbf{x})\rangle_L
\equiv\frac{1}{V_L}\int_{V_L} \zeta (\mathbf{x}) d^3\mathbf{x}=0,
\end{equation}
where $ V_L $ is the  comoving three-dimensional volume of the  domain.

From equations (\ref{locala}) and (\ref{constr}), it follows that
\begin{equation}
\label{backgra}a(t) = e^{\frac{1}{V_L}\int_{V_L} \ln\tilde{a} d^3\mathbf{x}}
\end{equation}
and
\begin{equation}
\label{zetagl}\zeta (\mathbf{x}) =\ln \tilde{a} - \frac{1}{V_L}
\int_{V_L} \ln\tilde{a} d^3\mathbf{x} .
\end{equation}

If the considered domain is infinite, one can use the Fourier
integral transformation
\begin{eqnarray}
\label{Fourie1}\zeta (\mathbf{x}) &=& \int \frac{d^3k}{(2\pi)^3}
\zeta_\mathbf{k}e^{i\mathbf{k}\mathbf{x}}, \\
\label{Fourie2}\zeta _\mathbf{k}&=&\int \zeta(\mathbf{x})
e^{-i\mathbf{k} \mathbf{x}}d^3x.
\end{eqnarray}

Assuming statistical homogeneity, the $n$-point correlators of
$\zeta_\mathbf{k}$ are of the form
\begin{eqnarray}
\label{2point}\langle \zeta_ {\mathbf{k}_1}\zeta_ {\mathbf{k}_2}\rangle
&=&(2\pi)^3\delta^{(3)}(\mathbf{k}_1+\mathbf{k}_2)P_\zeta(\mathbf{k}_1), \\
\label{3point}\langle\zeta _ {\mathbf{k}_1}\zeta_{\mathbf{k}_2}
\zeta_ {\mathbf{k}_3} \rangle &=&(2\pi)^3\delta^{(3)}
(\mathbf{k}_1+\mathbf{k}_2
+\mathbf{k}_3) B_\zeta(\mathbf{k}_1,\mathbf{k}_2, \mathbf{k}_3), \\
\label{4point}\langle\zeta_ {\mathbf{k}_1}\zeta _ {\mathbf{k}_2}
\zeta _ {\mathbf{k}_3} \zeta _ {\mathbf{k}_4} \rangle
&=&(2\pi)^3\delta^{(3)} (\mathbf{k} _1 +\mathbf{k}_2
+\mathbf{k}_3+\mathbf{k}_4) T_\zeta(\mathbf{k}_1), \mathbf{k}_2,
\mathbf{k}_3, \mathbf{k}_4).
\end{eqnarray}
Bispectrum  $B_\zeta(\mathbf{k}_1,\mathbf{k}_2, \mathbf{k}_3)$ and
trispektrum $T_\zeta(\mathbf{k}_1, \mathbf{k}_2, \mathbf{k}_3,
\mathbf{k}_4)$ are usually parameterized in terms of the spectrum
$P_\zeta(\mathbf{k})$ and the non-linear parameteres $f_{NL}$,
$\tau_{NL}$, $g_{NL}$  by
\begin{eqnarray}
\label{bs} B_\zeta(\mathbf{k}_1,\mathbf{k}_2,\mathbf{k}_3)
&=&\frac{6}{5}f_{NL}(\mathbf{k}_1,\mathbf{k}_2,\mathbf{k}_3)
\left(P_\zeta(\mathbf{k}_1\!)P_\zeta(\mathbf{k}_2\!) +
P_\zeta(\mathbf{k}_2\!)P_\zeta(\mathbf{k}_3\!)
+P_\zeta(\mathbf{k}_3\!)P_\zeta(\mathbf{k}_1\!) \right), \\
T_\zeta(\mathbf{k}_1,\mathbf{k}_2,\mathbf{k}_3,\mathbf{k}_4\!)
&=&\frac{1}{2}\tau_{NL}
(\mathbf{k}_1,\mathbf{k}_2,\mathbf{k}_3,\mathbf{k}_4\!)
\left(P_\zeta(\mathbf{k}_1\!)P_\zeta(\mathbf{k}_2\!)
P_\zeta(\mathbf{k}_1+\mathbf{k}_4\!) + cyclic~ permutations
\right) \nonumber \\
\label{trs}&+& \frac{54}{25}g_{NL}
(\!\mathbf{k}_1,\mathbf{k}_2,\mathbf{k}_3,\mathbf{k}_4\!)
\left(P_\zeta(\!\mathbf{k}_1\!)P_\zeta(\!\mathbf{k}_2\!)
P_\zeta(\!\mathbf{k}_3\!) + cyclic~ permutations \right)\! .
\end{eqnarray}
In the case of a finite domain, it is necessary to introduce
modifications of the equations (\ref{2point})-(\ref{4point}).

On a sufficiently large scale, it is very convenient to use the ansatz
\begin{equation}
\label{zetaanzatz}\zeta (\mathbf{x}) = \zeta_G (\mathbf{x})
+\frac{3}{5}f_{NL}^0\left(\zeta_G^2 (\mathbf{x})-\langle\zeta
_G^2\rangle\right)+\left(\frac{3}{5}\right)^2g_{NL}^0\left(\zeta_G^3
(\mathbf{x})-\langle\zeta _G^3\rangle \right) +... ,
\end{equation}
where the auxiliary quantity $ \zeta_G (\mathbf{x}) $ is a
Gaussian random variable and the non-linearity parameters
$f_{NL}^0$, $g_{NL}^0$ are some dimensionless constants. This
expression can be treated as $ \delta N $-formula for a simple
inflationary model with one single scalar field $ \varphi $
\begin{equation}
\zeta (\mathbf{x}) = \sum_{n=1}^\infty \frac{N^{(n)}
(\varphi_0(t_*))}{n!}\delta\varphi^n(t_*,\mathbf{x}),
\end{equation}
where $\varphi(t)$ is the background scalar field, $ t_ * $ is
some moment immediately after horizon crossing and the scalar
field perturbation $\delta\varphi^n(t_*,\mathbf{x}) = \zeta_G
(\mathbf{x})/ N '$ is assumed Gaussian.

Let us consider the patch $\Omega_S$ with  comoving volume $ V_S
$, which is located within the treated above domain. In this
patch, one can define the local comoving curvature perturbation,
denoted as $ \zeta_S $. Analogically  to equation (\ref{zetagl}),
this quantity is given by
\begin{equation}
\label{zetaloc}\zeta_S (\mathbf{x}) =\ln \tilde{a} -
\frac{1}{V_S}\int_{V_S} \ln\tilde{a} d^3\mathbf{x} .
\end{equation}

Since the volume averaging does not change the background scale
factor ($\langle a(t)\rangle_S=a(t)$), one can obtain
\begin{equation}
\zeta_S (\mathbf{x}) =\ln \left(\frac{\tilde{a}}{a}\right) -
\frac{1}{V_S}\int_{V_S} \ln \left(\frac{\tilde{a}}{a}\right)
d^3\mathbf{x} =\zeta(\mathbf{x}) - \frac{1}{V_S}\int_{V_S}
\zeta(\mathbf{x}) d^3\mathbf{x}.
\end{equation}
This equation can be rewritten as  \cite{1303.3549}
\begin{equation}
\label{zetaSzeta}\zeta_S (\mathbf{x})  =\zeta(\mathbf{x}) -
\langle \zeta\rangle_S.
\end{equation}

The equation (\ref{zetaSzeta}) shows that the relationship between
$\zeta_S$ and  $\zeta$ is non-local. To calculate the value of the
quantity $ \zeta_S $ at one point, it is necessary to know
$\zeta(\mathbf{x})$  in the whole patch at the same moment of
time. This non-locality is essential if the region  $\Omega_S$  is
not fixed, i.e., if $\Omega_S=\Omega_S(\mathbf{x})$. In this case,
to find $\zeta_S (\mathbf{x})$ at all points of
$\Omega_S(\mathbf{x}_0)$, it is necessary to know the comoving
curvature perturbation $\zeta(\mathbf{x})$ at all points of some
larger patch which includes $\Omega_S(\mathbf{x}_0)$.

\section{Mapping between $\zeta$ and $\zeta_S$ in the momentum space.}
\label{ch3}

The procedure of spatial averaging plays an important role in the
decomposition of the quantity $\tilde{a}(\mathbf{x},t) $ on the
homogeneous background and small inhomogeneous perturbation. Here,
we consider the relations in Fourier space, which follows from the
equation (\ref{zetaSzeta}).

We use the fact that the operation of spatial averaging is linear
and consider only one Fourier mode
\begin{equation}
\label{pw}\zeta(\mathbf{x})=\zeta (\mathbf{k})e^{i\mathbf{k}\mathbf{x}}.
\end{equation}
We denote the comoving coordinates of the patch as $ \mathbf{x} _0
$. For example, if the region has a spherical shape, then
$\mathbf{x}_0$ is the center of this sphere. We also introduce the
integration variable $\mathbf{x}'$, so that
$\mathbf{x}=\mathbf{x}_0+\mathbf{x}'$. The spatial averaging gives
\begin{equation}
\langle \zeta\rangle_S=\frac{1}{V_S}\int \zeta (\mathbf{k})
e^{i\mathbf{k}(\mathbf{x}_0+\mathbf{x}')}d(\mathbf{x}_0+\mathbf{x}')
= W(\mathbf{k})\zeta (\mathbf{k})e^{i\mathbf{k}\mathbf{x}_0},
\end{equation}
where
\begin{equation}
\label{Wk}W(\mathbf{k})=\frac{1}{V_S}\int e^{i\mathbf{k}
\mathbf{x}'} d\mathbf{x}'.
\end{equation}

Thus, the spatial averaging of the function (\ref{pw}) gives a
plane wave field with the same wave vector, but with a different
amplitude. We obtain the known results \cite{1303.3549}
\begin{equation}
\left[\langle \zeta\rangle_S \right]_\mathbf{k} = W(\mathbf{k})
\zeta_\mathbf{k}
\end{equation}
and
\begin{equation}
\label{zetak}\left[\zeta_S \right]_\mathbf{k} =(1-W(\mathbf{k}))
\zeta_\mathbf{k}.
\end{equation}
The equation (\ref{zetak}) is exact regardless of the form of the
quantity $\zeta (\mathbf{x})$. Although this equation is known,
its consequences for cosmological perturbation theory have not
been investigated.

In what follows, we assume that the considered patch is Hubble
sized at the present time. This choice allows us to get some
physical consequences of equation (\ref{zetak}).

Let's denote the characteristic comoving size of the patch as
$x_S$. The observer can confidently separate the perturbation from
the background if the perturbation experienced several
oscillations in the patch. This is carried out for the Fourier
modes with wave vector $\mathbf{k}$ satisfying the inequality
$|\mathbf{k}|x_S\gg 2\pi$. Such  perturbations we call short
wavelength ones. The long wavelength Fourier modes satisfy the
opposite condition $|\mathbf{k}|x_S\ll 2\pi$.

For short wavelength perturbations, the equations (\ref{Wk}),
(\ref{zetak}) give
\begin{equation}
\label{largek}\left[\zeta_S \right]_{\mathbf{k}} \Big|_{k\gg k_S}
= \left[\zeta \right]_{\mathbf{k}}\Big|_{k\gg k_S} ,
\end{equation}
where  $ k = |\mathbf{k}| $ and  $ k_S = 2\pi/x_S $. This equation
show that the difference between $\left[\zeta_S
\right]_{\mathbf{k}}$ and $\left[\zeta \right]_{\mathbf{k}}$ can
be neglected on a small scale.

Expanding the exponential function in a Taylor series, we get for
long wavelength  perturbation
\begin{equation}
\label{smallk}\left[\zeta_S \right]_{\mathbf{k}}\Big|_{k\ll k_S}
\approx \frac{1}{2}\langle(\mathbf{k}\mathbf{x}')^2
\rangle_S\left[\zeta \right]_{\mathbf{k}}\Big|_{k\ll k_S}\propto
k^2\left[\zeta \right]_{\mathbf{k}}\Big|_{k\ll k_S}.
\end{equation}
One can see that quantity  $\left[\zeta_S \right]_{\mathbf{k}}$ is
suppressed on a large scale. This is the  consequence of the fact
that the observer can not separate surely the long-wavelength
perturbation from the background. Perhaps, the equation
(\ref{smallk}) will allow to solve still existing problem of
infrared divergences in the calculation of correlation functions
at loop level (see \cite{1306.4461} for a recent review of this
topic), although the reformulation of the perturbation theory in
terms of locally defined variables is difficult due the
non-locality of the equation (\ref{zetaSzeta}).

The form of the function $W(\mathbf{k})$ is of interest over a
wide range of scales. Assuming that the observable patch can be
approximated as a sphere of radius $ x_S $, a simple calculation
gives
\begin{equation}
\label{WkSphere}W(\mathbf{k})=3\left(\frac{\sin (kx_S)}{(kx_S)^3}
- \frac{\cos (kx_S)}{(kx_S)^2} \right).
\end{equation}
Equations (\ref{largek}) and (\ref{smallk}) are main results of
this work.

Since the quantity $\zeta_S(\mathbf{x})$ is statistically
homogeneous, there are an expressions which are analogous to the
equations (\ref{2point})- (\ref{trs}). In particular, we have
\begin{equation}
\label{bsS} B_{\zeta_S}\!(\mathbf{k}_1,\mathbf{k}_2,\mathbf{k}_3)
\!=\!\frac{6}{5}f_{NL}^{(S)} \!(\mathbf{k}_1, \mathbf{k}_2,\mathbf{k}_3)
\left(P_{\zeta_S}\!(\mathbf{k}_1\!)\!P_{\zeta_S}\!(\mathbf{k}_2\!) +
P_{\zeta_S}\!(\mathbf{k}_2\!)\!P_{\zeta_S}\!(\mathbf{k}_3\!)
+P_{\zeta_S}\!(\mathbf{k}_3\!)\!P_{\zeta_S}\!(\mathbf{k}_1\!) \right).
\end{equation}

For equal-$k$ ($k_1=k_2=k_3\equiv k$), the equations
(\ref{bs}),(\ref{bsS}) gives
\begin{equation}
f_{NL}^{(S)} (\mathbf{k}_1, \mathbf{k}_2, \mathbf{k}_3) =
\frac{f_{NL}}{1-W(k)}.
\end{equation}
In the squeezed limit ($k_1 \approx k_2\equiv k\gg k_3$), the
mutual relations of $f_{NL}^{(S)}$ and $f_{NL}$ depends on the
scale. If $k_1$, $k_2$ are infrared and the primordial spectrum is
scale-invariant,  then
\begin{equation}
f_{NL}^{(S)}  (\mathbf{k}_1, \mathbf{k}_2, \mathbf{k}_3) \approx
f_{NL } (\mathbf{k}_1, \mathbf{k}_2, \mathbf{k}_3)
\frac{2}{1-W(k)} \frac{k}{k_3}.
\end{equation}
In both cases, we obtain
\begin{equation}
f_{NL}^{(S)} (\mathbf{k}_1, \mathbf{k}_2, \mathbf{k}_3) \gg f_{NL}
(\mathbf{k}_1, \mathbf{k}_2, \mathbf{k}_3).
\end{equation}
Formally, the non-linearity parameter $f_{NL}^{(S)}$ is enhanced
on a large scale due the suppression of $\left[\zeta_S \right]
_\mathbf{k}$.

\section{Examples.}
\label{ch4}

The equation (\ref{largek}) indicates that, on a small scale, the
local comoving curvature perturbation $ \zeta_S $ has the same
statistical properties as the quantity $ \zeta $. The only
question is whether it is possible to recover the global
statistical properties of $ \zeta_S $ using the observed data
sample of $\left[ \zeta_S \right]_\mathbf{k}$. In other words, are
there any reasons for which the local observer  can fail to notice
the coincidence of statistical properties of $\left[ \zeta_S
\right]_\mathbf{k}$ and $\left[ \zeta \right]_\mathbf{k}$ on a
small scale?

At first, let us consider how to implement the equation
(\ref{largek}). For simplicity, we use the ansatz
(\ref{zetaanzatz}), in which we assume that only $f_{NL}^0\neq 0$.
The ansatz (\ref{zetaanzatz}) is reduced to
\begin{equation}
\label{zetafnl}\zeta (\mathbf{x}) = \zeta_G (\mathbf{x})
+\frac{3}{5}f_{NL}^0\left(\zeta_G^2 (\mathbf{x})-\langle\zeta
_G^2\rangle\right)
\end{equation}
and leads to the equation
\begin{equation}
\label{zetaSfnl}\zeta_S (\mathbf{x}) =  \zeta_G (\mathbf{x}) -
\langle \zeta_G\rangle_S +\frac{3}{5}f_{NL}^0\left(\zeta_G^2
(\mathbf{x}) -\langle\zeta_G^2\rangle_S \right).
\end{equation}

In momentum space it gives
\begin{equation}
\label{zetaSfnlk}\left[ \zeta_S \right]_\mathbf{k} =\left[ \zeta_G
- \langle \zeta_G\rangle_S \right]_\mathbf{k} +\frac{3}{5}
f_{NL}^0 \left[\zeta_G^2 -\langle\zeta_G^2 \rangle_S
\right]_\mathbf{k}.
\end{equation}
The  volume averaging over the Hubble patch cuts off short
wavelength perturbations, so that on a small scale
\begin{equation}
\label{zetaSk1}\left[ \zeta_S \right]_\mathbf{k} =\left[ \zeta_G
\right]_\mathbf{k} +\frac{3}{5}f_{NL}^0\left[\zeta_G^2
\right]_\mathbf{k} .
\end{equation}
It coincides with the result (\ref{largek}).

Observable patch is differs from the large domain by size, and the
discrepancy between observed statistical properties of $\left[
\zeta_S \right]_\mathbf{k}$ and theoretical statistical properties
of $\left[ \zeta \right]_\mathbf{k}$ may be expected to be due to
the influence of infrared perturbations. Let us examine this issue
in some detail.

Let's the realization of the auxiliary quantity $ \zeta_G $ has
the form
\begin{equation}
\label{twomode}\zeta_G=\zeta_1+\zeta_2,
\end{equation}
where $\zeta_1=\zeta_{\mathbf{k}_1}e^{\mathbf{k}_1 \mathbf{x}}$,
$\zeta_2=\zeta_{\mathbf{k}_2}e^{\mathbf{k}_2\mathbf{x}}$ and $k_1
\ll k_S$ (long wave), $k_2 \gg k_S$ (short wave). Then the
equation (\ref{zetaSfnlk}) yelds
\begin{equation}
\label{zetatm}\left[\zeta_S \right]_{\mathbf{k}_2} = \left[
\zeta_G \right]_{\mathbf{k}_2} +\frac{3}{5}f_{NL}^0\left[\zeta_G^2
-\langle\zeta_G^2\rangle_S\right]_{\mathbf{k}_2} .
\end{equation}

In the coordinate space, we obtain from equations (\ref{twomode}),
(\ref{zetatm})
\begin{equation}
\zeta_G^2  -\langle\zeta_G^2\rangle_S =   \left(\zeta_{2}^2 -
\langle\zeta_{2}^2 \rangle_S \right)+ 2\zeta_{1} \left( \zeta_{2}
- \langle\zeta_{2}\rangle_S\right).
\end{equation}

In the momentum space it gives
\begin{equation}
\left[\zeta_G^2  -\langle\zeta_G^2\rangle_S \right]
_{\mathbf{k}_2} =   \left[ \left(\zeta_{2}^2 - \langle\zeta_{2}^2
\rangle_S \right)+ 2\zeta_{1} \left( \zeta_{2}  -
\langle\zeta_{2}\rangle_S\right)\right]_{\mathbf{k}_1}=0.
\end{equation}

Using this result one can obtain
\begin{equation}
\left[\zeta_S \right]_{\mathbf{k}_2} = \left[ \zeta_G
\right]_{\mathbf{k}_2},
\end{equation}
i. e., the long wavelength Fourier mode with wave vector
$\mathbf{k}_1$ does not affect the mode with wave vector
$\mathbf{k}_2$. It is clear also from consideration of the
convolution theorem
\begin{equation}
\label{conv}\left[\zeta_G ^2 \right]_{\mathbf{k}}=\int
\left[\zeta_G \right]_{\mathbf{q}}\left[\zeta_G \right]
_{\mathbf{k}-\mathbf{q}}\frac{d^3\mathbf{q}}{(2\pi)^3}.
\end{equation}

If the quantity $\zeta_G$ has two modes with wave vectors
$\mathbf{k}_1$ and $\mathbf{k}_2$, then the quantity $\zeta_G^2$
contain modes  with wave vectors $2\mathbf{k}_1$, $2\mathbf{k}_2$,
$\mathbf{k}_1 + \mathbf{k}_2$ only.

To gain influence on mode with the wave vector $\mathbf{k}_2$, the
expansion of $\zeta_G$  should include à mode with $\mathbf{k}_3 =
\mathbf{k}_2-\mathbf{k}_1$. However, just like the Fourier mode
with $\mathbf{k}_1$ is indistinguishable from the background, the
modes with $\mathbf{k}_2$ and $\mathbf{k}_3$ are indistinguishable
between themselves. In other words, infrared perturbations do not
contribute to the  correlation functions of physically
distinguishable modes.

\section{A landscape Universe.}
\label{ch5}

Recently, in the papers \cite{1301.3128},\cite{1303.3549}, it was
made several claims, which seem to contradict the equation
(\ref{largek}). However, the authors of
\cite{1301.3128},\cite{1303.3549} have used several approximations
which require very careful handling when working in Fourier space.
On the other hand, the parameters of ansatz (\ref{zetaanzatz}) are
sensitive to the choice of the auxiliary quantity $\zeta_G$. This
can be verified by direct calculation in the simplest case of
anzatz (\ref{zetaSfnl}).

We introduce the notation
\begin{equation}
\label{GS}\zeta_{G,S}=\zeta_G-\bar{\zeta},
\end{equation}
where $\bar{\zeta}=\langle \zeta_G\rangle_S$. It is important that
the quantity $\tilde{\zeta}$ can be considered as a constant on a
small scale, i.e., one can assume
\begin{equation}
\label{kmean}\bar{\zeta} _{\mathbf{k}}\Big|_{k\gg k_S} =0.
\end{equation}

The equation (\ref{zetaSfnl}) and the equality
\begin{equation}
\label{eqconstr}\zeta_{G,S}^2-\langle\zeta_{G,S}^2 \rangle_S =
\zeta_G^2 - \langle\zeta_G^2\rangle-2\bar{\zeta}\zeta_{G,S}
\end{equation}
gives
\begin{equation}
\label{lanscape1}\zeta_S (\mathbf{x}) =\left(1 + \frac{6}{5}f_{NL}
\bar{\zeta}\right) \zeta_{G,S} +\frac{3}{5}f_{NL}^0
\left(\zeta_{G,S}^2 -\langle\zeta_{G,S}^2\rangle_S \right).
\end{equation}
This result is in full agreement with the result of the paper
\cite{1303.3549}. On a small scale, the equations (\ref{kmean}),
(\ref{lanscape1}) yields now
\begin{equation}
\left[ \zeta_S \right]_\mathbf{k}  =\left(1 + \frac{6}{5}f_{NL}^0
\bar{\zeta}\right) \left[\zeta_{G,S}\right]_\mathbf{k}
+\frac{3}{5}f_{NL}\left[\zeta_{G,S}^2\right]_\mathbf{k}.
\end{equation}

One can make the variable transformation
\begin{equation}
\label{chix}\chi_G=\left(1 + \frac{6}{5}f_{NL}^0 \bar{\zeta}
\right) \zeta_{G,S}.
\end{equation}
It gives \cite{1303.3549}
\begin{equation}
\label{zetaSk2}\left[ \zeta_S \right]_\mathbf{k} = \left[\chi_{G}
\right] _\mathbf{k} +\frac{3}{5}f_{NL}^{0(S)} \left[\chi_{G}^2
\right] _\mathbf{k},
\end{equation}
where
\begin{equation}
\label{fnl0s}f_{NL}^{0(S)}=\frac{f_{NL}^0}{\left(1 +
\frac{6}{5}f_{NL}^0 \bar{\zeta}\right)^2}.
\end{equation}
Consequently, both parameterizations (\ref{zetaSk2}) and
(\ref{zetaSk1}) are possible at once. This can easily be checked
on a small scale using the equalities
\begin{eqnarray}
\left[\zeta_{G,S}\right]_\mathbf{k} &=& \left[\zeta_{G}
\right]_\mathbf{k},\\
\label{GS2} \left[\zeta_{G,S}^2\right]_\mathbf{k} &=&
\left[\zeta_{G}^2\right]_\mathbf{k} -2\bar{\zeta} \left[\zeta_{G}
\right]_\mathbf{k}.
\end{eqnarray}
which follows from the equations (\ref{GS}) and (\ref{eqconstr})
correspondingly.  The parameterizations (\ref{zetaSk1}) and
(\ref{zetaSk2}) yields to the same numerical values for the
Fourier components of the local comoving curvature perturbation,
as it should be.

It is significant that  the convolution theorem (\ref{conv})
allows us to rewrite the equation (\ref{GS2}) as
\begin{equation}
\left[\zeta_{G,S}^2\right]_\mathbf{k} = \int \left[\zeta_G
\right]_{\mathbf{q}}\left[\zeta_G \right]_{\mathbf{k}
-\mathbf{q}}\frac{d^3\mathbf{q}}{(2\pi)^3} -2\bar{\zeta}
\left[\zeta_{G} \right]_\mathbf{k}.
\end{equation}
This expression differs from the approximate one used in
the papers  \cite{1301.3128}, \cite{1303.3549}.

Equations (\ref{chix}), (\ref{GS2}) and (\ref{fnl0s}) indicate
that the quantities $\chi_G$ and $f_{NL}^{0(S)}$ are influenced by
the same random parameter $\bar{\zeta}$. Though, the associated
uncertainty of parametrization did not affect the numerical value
of the local comoving curvature perturbation. One can see that
some arbitrariness in definition of the auxiliary  function
$\chi_G$ is compensated by a corresponding change of the
non-linearity parameter $f_{NL}^{0(S)}$ without affecting  $\left[
\zeta_S \right]_\mathbf{k}$ and its correlation functions. The
stochastic quantity $\bar{\zeta}$ also does not affect the
calculation of the parameter $f_{NL}^{(S)}$ defined by the
equation  (\ref{bsS}).

\section{Conclusions.}
\label{ch6}

We have studied the relationship between locally and globally
defined comoving curvature perturbations within a
quasi-homogeneous domain of the Universe. It is shown that, on
scales larger than the size of the observed patch, the Fourier
components of the locally defined comoving curvature perturbation
are suppressed. It is shown also  that the statistical properties
of local and global comoving curvature perturbations are coincide
on a small scale. Several examples are discussed in detail. We
have shown that, in the simplest cases,  the  long wavelength
perturbations do not contribute to the bispectrum and trispektrum
on a sub-Hubble scale.

\end{document}